# Resolution enhancement of NMR by decoupling with low-rank Hankel model

Tianyu Qiu[a,b], Amir Jahangiri [b], Xiao Han[c], Dmitry Lesovoy[d], Tatiana Agback[e], Peter Agback[e], Adnane Achour[c], Xiaobo Qu *[a], and Vladislav Orekhov *[b]

Nuclear magnetic resonance (NMR) spectroscopy has become a formidable tool for biochemistry and medicine. Although J-coupling carries essential structural information it may also limit the spectral resolution. Homonuclear decoupling remains a challenging problem. In this work, we introduce a new approach that uses a specific coupling value as prior knowledge, and Hankel property of exponential NMR signal to achieve the broadband heteronuclear decoupling using the low-rank method. Our results on synthetic and realistic HMQC spectra demonstrate that the proposed method not only effectively enhances resolution by decoupling, but also maintains sensitivity and suppresses spectral artefacts. The approach can be combined with the non-uniform sampling, which means that the resolution can be further improved without any extra acquisition time.

## Introduction

Nuclear magnetic resonance (NMR) spectroscopy is a widely used technique in chemistry[1], biology[2] and medicine[3]. Resolution enhancement plays an important role in NMR since it determines the quality of the quantitative and qualitative analysis. The improvement of hardware, such as higher magnetic fields, has significantly enhanced resolution[4,5]. Nevertheless, there are still two main problems that limits spectral resolution.

According to signal processing theory, resolution enhancement requires long acquisition time, *i.e.* more measured data points in the time domain. In multidimensional NMR experiments, this forces the use of very long total measurement time, which is proportional to the number of points for indirect spectral dimensions. However, the appearance of non-uniform sampling (NUS) and reconstruction methods, such as maximum entropy[6,7], compressed sensing (CS)[8], multi-dimensional decomposition(MDD)[9,10], low-rank Hankel method (LR) and more recently deep learning-based techniques[11-13], have greatly alleviated this problem.

The homo-nuclear J-coupling causes signal splitting and thus represents another reason for line-broadening and loss of resolution. The decoupling can be achieved in several ways, including the use of the pure shift approach[14,15], constant time evolution[16,17], bilinear rotational decoupling[18,19], etc.

The mechanism of the J-coupling is well understood and the coupling values are known[20,21]. This information can be exploited to perform decoupling by software deconvolution also known as virtual decoupling (VD)[11,13,22-24]. Decoupling and reconstruction of spectrum from NUS data can be therefore combined and solved by one single method. Furthermore, it was noted that VD is likely to improve NUS reconstruction, because it reduces the number of individual peaks in the spectrum, which can have related implications for different reconstruction algorithms. Thus, in compressed sensing [8], VD increases sparseness of the spectrum. Similarly, the Low-Rank (LR) reconstruction [25-28], which is based on the low-rank Hankel property of the time domain free induction decay (FID) NMR signal, benefits from the VD, because the splitting caused by J-coupling increases the number of peaks and consequentially the rank. This requires an increase of NUS levels or even, when the Hankel matrix is not low-rank anymore, may corrupt the reconstructed spectrum.

In this work, we used a specific coupling value as prior knowledge so that the FID can be reconstructed and decoupled simultaneously. Since the decoupling reduces the number of peaks in the spectrum, the NUS fraction can be further decreased.

## Method

FID signal is expressed as [28-30]:

$$x_0(t) = \sum_{r=1}^{R} a_r e^{(j2\pi f_r - \tau_r)t}, \quad (1)$$

where $a_r$, $f_r$ and $\tau_r$ denote the complex amplitude, the central frequency and the damping factor, respectively, and the summation goes over all $R$ peaks in the spectrum.

For a J-coupled two-spin systems, the FID signal is written as:

$$x_c(t) = x_0(t)c(t), \quad (2)$$

where $c(t) = \cos \pi J t$ [24] and J represents the coupling value.

The proposed Low-Rank Decoupling (LRD) method is expressed as:

$$\min_{\boldsymbol{x}} \|\mathcal{R}\boldsymbol{x}\|_* + \frac{\lambda}{2}\|\boldsymbol{y} - \mathcal{U}\boldsymbol{C}\boldsymbol{x}\|_2^2, \quad (3)$$

where vector $\boldsymbol{x}$ stands for the variables that needs to be determined. Vector $\boldsymbol{y}$ represents the measurement with coupling. Matrix $\boldsymbol{C}$ is defined as $diag(\boldsymbol{c})$ which denotes the finite discrete form of Equation (2). Operator $\mathcal{R}$ transforms a vector into a Hankel matrix. Operator $\mathcal{U}$ represents the NUS schedule. The entire algorithm is presented and summarized in the supplemental material (SI) section.

## Results

We used simulated and experimental spectra to verify the performance of the LRD methodology. The one-bond couplings occur between adjacent $^{13}$C atoms, *e.g.* in proteins $C_\alpha$-$C_\beta$ backbone pairs or between methyl carbons, and their adjacent carbons. It should also be noted that we assumed within the frame of the present study (*i.e.* in Eqs. 2,3) the typical value for these coupling values as $^1J_{CC} = 35$ Hz [31].

The proposed method was compared with conventional decoupling using the iteratively reweigthed least square algorithm for comopressed sensing (CS-IRLS) [24] algorithm implemented in the mddnmr software[32]. The compared method utilizes the same assumption about the decoupling value, but constrains the sparsity in Fourier spectra. For experimental signals, a spectrum decoupled by constant time (CT) evolution sequence, which is very commonly used in most applications[16], has been added for comparison.

To ensure a fair comparison, merely the decoupling of fully-sampled spectra is presented in the main text. The decoupling of NUS signals is presented in SI section. In SI, the reconstruction from 40% NUS data has been shown, illustrating the clear possibility to combine LRD with NUS, resulting in a remarkable improvement of resolution without any necessary extra acquisition time.

**Simulation**

The results presented in Fig. 1 display a comparison between LRD and CS on a synthetic spectrum. Both methods successfully decouple the spectrum as shown in Fig. 1(b). While CS-IRLS offers a spectrum with a perfect baseline, it also over-sharpens the resonances. Furthermore, it may also weaken the low intensity peaks (note for example the peak marked by arrow in Fig. 1(c)). In contrast, the LRD method performs well, preserving the intensity and providing a comparatively better line shape (as marked by arrows in Fig. 1(d)).

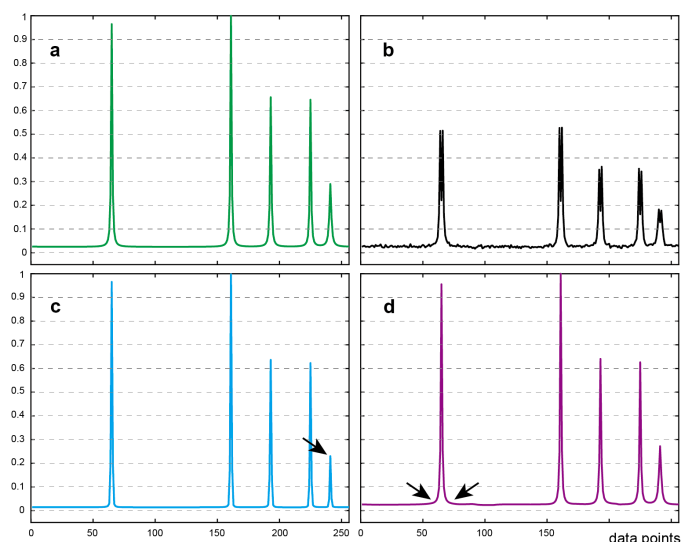

**Fig. 1**. **The virtual decoupling in a synthetic spectrum containing five peaks. (a)** is the reference fully-sampled spectrum without J-coupling. **(b)** the spectrum with J=35 Hz. **(c)** and **(d)** are decoupled spectra by CS-IRLS and by the LRD method proposed within the present study, respectively. Arrows points to the peaks mentioned in the text.

**Experimental data**

In this part, a 2D HMQC spectrum of 44kDa fragment of the mucosa-associated lymphoma translocation protein 1 (MALT1[Casp-IgL3]$_{338-719}$) is used as an example[33]. The details of all the performed experiments including samples, equipment and acquisition parameters, are presented in the SI section.

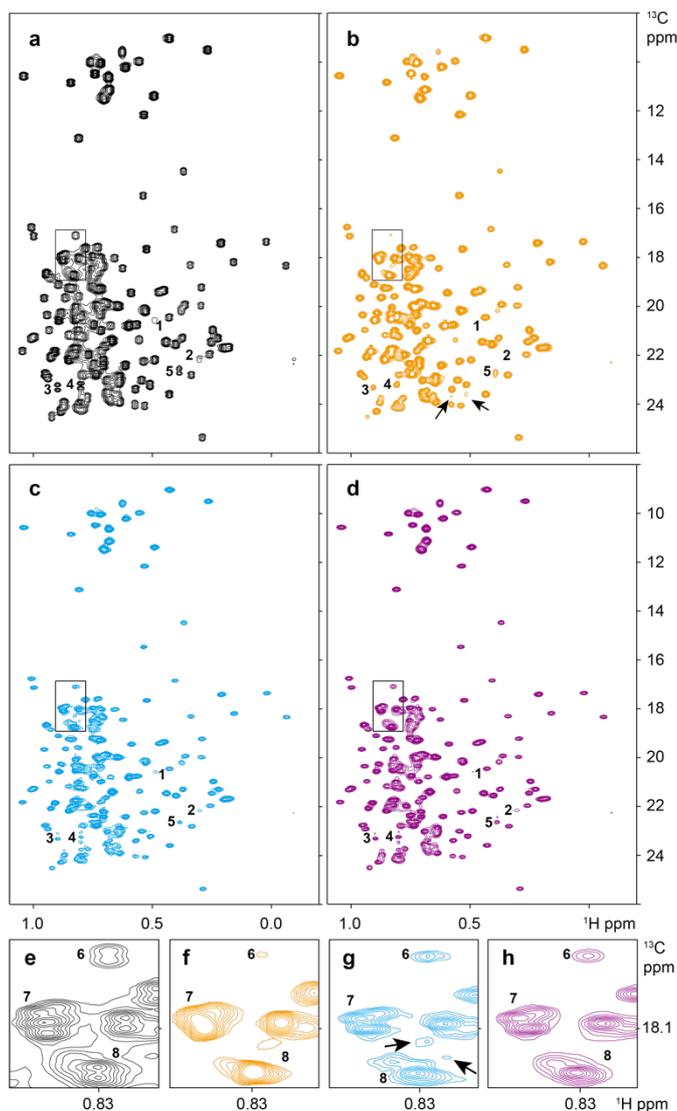

**Fig. 2. Decoupling of a 2D $^1$H-$^{13}$C HMQC spectrum of MALT1. (a)** is the J-coupled spectrum. **(b)** is the spectrum obtained using the CT evolution sequence. **(c)** and **(d)** are virtually decoupled by CS-IRLS and by the proposed method LRD, respectively. Regions marked by black rectangles in (a)-(d) are enlarged in **(e)-(h)**, respectively. Peaks and spectral artefacts discussed in the text are indicated by numbers and black arrows, respectively.

Figure 2 displays different decoupling schemes in 2D $^1$H-$^{13}$C HMQC spectrum of MALT1. Although all three tested methodologies decouple the spectrum successfully, clearly noticeable differences in resolution, sensitivity and artefacts can be identified.

Compared to the J-couple spectrum (Fig. 2a), the constant time (CT) evolution offers a spectrum (Fig. 2b) with higher resolution. However, CT may also result in significant sensitivity loss. Indeed, some peaks are clearly weakened, such as peak 6

(Fig. 2(e)-(h)). Peaks with low intensity (such as peaks 1 and 2) even disappear or are completely covered by noise (Fig. 2(e)-(h)). Both the CS-IRLS and LRD methods decoupled spectra show better resolution than their CT counterpart due to longer acquisition time allowed by the these two VD approach. As a result, several resolved individual peaks emerged as for example peak groups 7 and 8. These peaks were not discernible in the spectrum decoupled by constant time evolution. In CT experiments, the resolution can be improved only to the expense of further significant loss of sensitivity.

For both CS-IRLS and LRD, some artefacts such as peaks 3, 4 and 5, are caused by significant deviations of the actual coupling from 35 Hz value used for reconstruction. This problem can be alleviated in some applications. For example, in HNCA experiments, coupling variations are usually small[24]. However, as shown in Fig. 2(g), there are some other unignorable artefacts in the CS-IRLS spectrum marked by black arrows. The proposed LRD method seemingly provides a cleaner spectrum, which helps to avoid to introducing ambiguity in quantitative and qualitative analyses.

## Conclusions

We present here a new decoupling methodology, named LRD, which is based on low-rank Hankel model and the introduction of specific coupling values for one-bond coupling. The coupling between adjacent $^{13}C$ atoms was taken as an example for the validation of our approach and for comparison with other already established methods. Our obtained decoupling results, on both synthetic and experimental spectra, demonstrate that the LRD method is capable of decoupling, offering higher resolution and significantly cleaner spectra. The presented approach provides a new tool for broadband homonuclear decoupling.

## Author Contributions

Proposed model and designed numerical experiments: T. Qiu, V. Orekhov. Performed numerical experiment: T. Qiu, A. Jahangiri, and D. Lesovoy. Analysed data: T. Qiu, V. Orekhov. Contributed samples: X. Han, T. Agback, P. Agback, A. Achour. Wrote the paper: T. Qiu, V. Orekhov, X. Qu.

## Acknowledgements

This work was supported, in part, by the National Natural Science Foundation of China (grants 62122064, 61871341, and 82071913), Xiamen University Nanqiang Outstanding Talents Program, and Chinese Scholarship Council; The Swedish Foundation for Strategic Research grant ITM17-0218 to T.A and P.A., grant RSF 19-74-30014 to D.L., Swedish Cancer Society grant 21 1605 Pj01H to A.A., and the Swedish Research Council grants 2021-05061 to A.A. and 2019-03561 to V.O. This study used NMRbox: National Center for Biomolecular NMR Data Processing and Analysis, a Biomedical Technology Research Resource (BTRR), which is supported by NIH grant P41GM111135 (NIGMS).

# Supplementary Information

## Algorithm

The model of the proposed LRD method is defined as:

$$\min_{x}\|\mathcal{R}x\|_* + \frac{\lambda}{2}\|y - \mathbf{P}\mathbf{C}x\|_2^2, \quad (A1)$$

where vector $\mathbf{x} \in \mathbb{C}^{N \times 1}$ stands for the variable that needs to be determined. Vector $y \in \mathbb{C}^{M \times 1}$ is the measurement data with coupling. Matrix $\mathbf{C}$ is defined as $diag(\mathbf{c})$ which denotes the finite discrete form of Eq. (2) presented in the main text of the manuscript. Operator $\mathcal{R}$ transforms a vector into a Hankel matrix. Matrix $\mathbf{P} \in \mathbb{R}^{M \times N}$ (M≤N) represents the NUS schedule. Symbol $\|\cdot\|_*$ denotes the nuclear norm defined as the sum of singular values. The regularization parameter $\lambda$ balances the nuclear norm and consistency.

By introducing two variables $\mathbf{Z}$ and $\mathbf{D}$, the augmented Lagrangian formulation of Eq. (A1) is written as:

$$\min_{\mathbf{x},\mathbf{Z}} \max_{\mathbf{D}} \|\mathbf{Z}\|_* + \frac{\beta}{2}\|\mathcal{R}\mathbf{x} - \mathbf{Z}\|_F^2 + \langle \mathbf{D}, \mathcal{R}\mathbf{x} - \mathbf{Z}\rangle + \frac{\lambda}{2}\|y - \mathbf{P}\mathbf{C}x\|_2^2, \quad (A2)$$

where $\langle \cdot, \cdot \rangle$ denotes the inner product in complex matrices. It is defined as $\langle \mathbf{A}, \mathbf{B}\rangle = \Re(tr(\overline{\mathbf{A}}\mathbf{B}))$, where $\overline{\mathbf{A}}$ represents the conjugation of $\mathbf{A}$, and the symbol $\Re$ denotes the real part[1].

Using alternating direction method of multipliers (ADMM)[2], the problem in Eq. (A2) is divided into the following three sub-problems:

$$\begin{cases} \min_{\mathbf{x}} \frac{\beta}{2}\left\|\mathcal{R}\mathbf{x} - \mathbf{Z} + \frac{\mathbf{D}}{\beta}\right\|_F^2 + \frac{\lambda}{2}\|y - \mathbf{P}\mathbf{C}x\|_2^2 \\ \min_{\mathbf{Z}} \|\mathbf{Z}\|_* + \frac{\beta}{2}\left\|\mathcal{R}\mathbf{x} - \mathbf{Z} + \frac{\mathbf{D}}{\beta}\right\|_F^2 \\ \mathbf{D} \leftarrow \mathbf{D} + \tau(\mathcal{R}\mathbf{x} - \mathbf{Z}) \end{cases}. \quad (A3)$$

The solution to Eq. (A3) is expressed as:

$$\begin{cases} \mathbf{x}_{k+1} = (\beta \mathcal{R}^*\mathcal{R} + \lambda \mathbf{C}^H \mathbf{P}^H \mathbf{P} \mathbf{C})^{-1}\left[\beta \mathcal{R}^*\left(\mathbf{Z}_k - \frac{\mathbf{D}_k}{\beta}\right) + \lambda \mathbf{C}^H \mathbf{P}^H y\right] \\ \mathbf{Z}_{k+1} = \mathcal{S}_{1/\beta}\left(\mathcal{R}\mathbf{x}_k + \frac{\mathbf{D}_k}{\beta}\right) \\ \mathbf{D}_{k+1} = \mathbf{D}_k + \tau(\mathcal{R}\mathbf{x}_{k+1} - \mathbf{Z}_{k+1}) \end{cases}, \quad (A4)$$

where the subscripts k and k+1 denote the iteration steps. The superscripts *H* and * denote conjugation transposed and adjoint operator, respectively. $\beta$ and $\tau$ are two parameters. $\mathcal{S}_{1/\beta}$ is a singular thresholding operator defined as $\mathcal{S}_{1/\beta}(\mathbf{X}) = \mathbf{U}diag(\{\sigma_r - 1/\beta\}_+)\mathbf{V}^H$, where matrix $\mathbf{X}$ is with singular value decomposition $\mathbf{X} = \mathbf{U}diag(\{\sigma_r\}_{r=1}^R)\mathbf{V}^H$ and $t_+ = max(0, t)$[3]. $\mathcal{R}^*$ denotes an operator that transforms a matrix into a vector by summarizing each skew diagonal. $\mathcal{R}^*\mathcal{R}$ is defined as an operator satisfying $\mathcal{R}^*\mathcal{R}\ \mathbf{x} = \mathbf{W}\mathbf{x}$, where $\mathbf{W}$ is a diagonal matrix whose main diagonal is the number of times that an element of $\mathbf{x}$ appears in a Hankel matrix[4].

The whole algorithm has been summarized as pseudo code in Table S1.

Table S1. Pseudo code of the LRD algorithm

| |
|---|
| Input: **y**, **C**, **P**, $\lambda$; Output: $\hat{\mathbf{x}}$ |
| Initialization: $k=1, \beta=1, \tau=1, k_{max}=2000, \mathbf{x}_1 = \mathbf{y}$ |
| 1)     While $k < k_{max}$ and $\|\mathbf{x}_{k+1} - \mathbf{x}_k\|_2 / \|\mathbf{x}_k\|_2 > 10^{-5}$, **do** <br> 2)     $\mathbf{x}_{k+1} = (\beta \mathcal{R}^* \mathcal{R} + \lambda \mathbf{C}^H \mathbf{P}^H \mathbf{PC})^{-1} [\beta \mathcal{R}^* (\mathbf{Z}_k - \mathbf{D}_k/\beta) + \lambda \mathbf{C}^H \mathbf{P}^H \mathbf{y}]$; <br> 3)     $\mathbf{Z}_{k+1} = \mathcal{S}_{1/\beta}(\mathcal{R}\mathbf{x}_k + \mathbf{D}_k/\beta)$; <br> 4)     $\mathbf{D}_{k+1} = \mathbf{D}_k + \tau(\mathcal{R}\mathbf{x}_{k+1} - \mathbf{Z}_{k+1})$; <br> 5)     $k \leftarrow k+1$; <br> 6)     **End while** |
| Output: $\hat{\mathbf{x}} = \mathbf{x}_{k+1}$ |

## Experiment for the 2D $^1$H- $^{13}$C HMQC spectrum

The sample used for the results presented in Fig. 2 was prepared as described before[5].

Fully-sampled methyl 2D $^1$H-$^{13}$C HMQC with 200 complex points in the $^{13}$C dimension (46.5 ms acquisition time) was acquired at 298K on a 900 MHz Bruker AVANCE III-HD spectrometer equipped with 3mm cryo-TCI probe. The directly detected dimension of the region of the full reference 2D spectrum (from 1.1 to -0.8 $^1$H ppm) was processed using the NMRPipe software[6], and imported in MATLAB and qMDD[7] for consecutive reconstruction and decoupling by LRD or CS-IRLS[8]. The NUS schedule satisfying Poisson gap[9] with 40% is generated along the $^{13}$C dimension. The parameter of Poisson distribution satisfies a sine-weighted function, changing as the location of sampled points.

The constant-time 2D $^1$H-$^{13}$C CT-HMQC (presented in Fig. 2(b) in the main text) with 102 complex points in the $^{13}$C dimension (22.5 ms acquisition time) was acquired at 298K on a 900 MHz Bruker AVANCE III-HD spectrometer equipped with a 3mm cryo-TCI probe.

## Decoupling from NUS data

The results of our study demonstrate that the proposed LRD method is capable of successfully decoupling fully-sampled spectrums. It is well established that NUS provides a reliable way to enhance resolution. Here, we introduced NUS to our model and verified the effects of this combination on synthetic and experimental spectra, demonstrating that the proposed methodology can significantly improve resolution without any extra acquisition time.

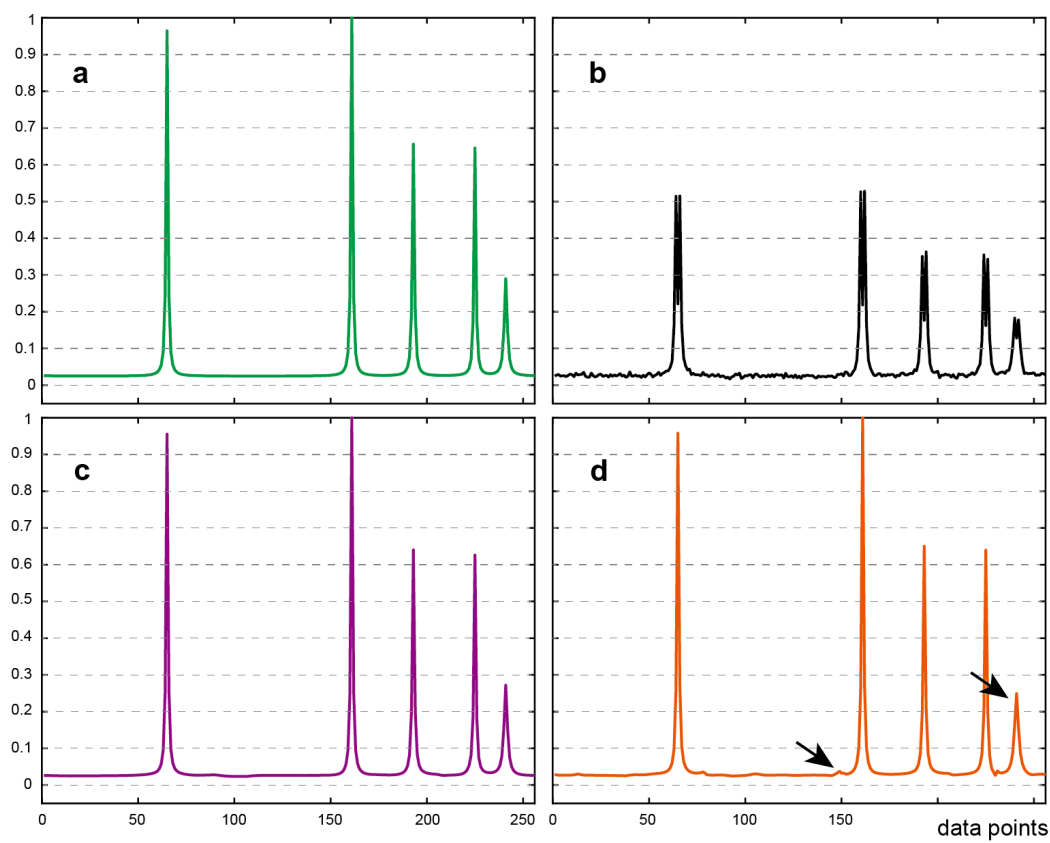

**Fig. S1. Decoupled synthetic spectra obtained by using the method presented within the present work**. **(a)** denotes the fully-sampled reference spectrum. **(b)** stands for the J-coupled spectrum with J=35 Hz. **(c)** and **(d)** are decoupled spectra from fully-sampled and 25% NUS data, respectively. Black arrows indicate low intensity artefacts at the baseline and the partly weakened lowest peak. It should be noted that a 1D NUS schedule satisfying the Poisson gap[9] was used. The standard deviation of the Gaussian noise in (b) was 0.005.

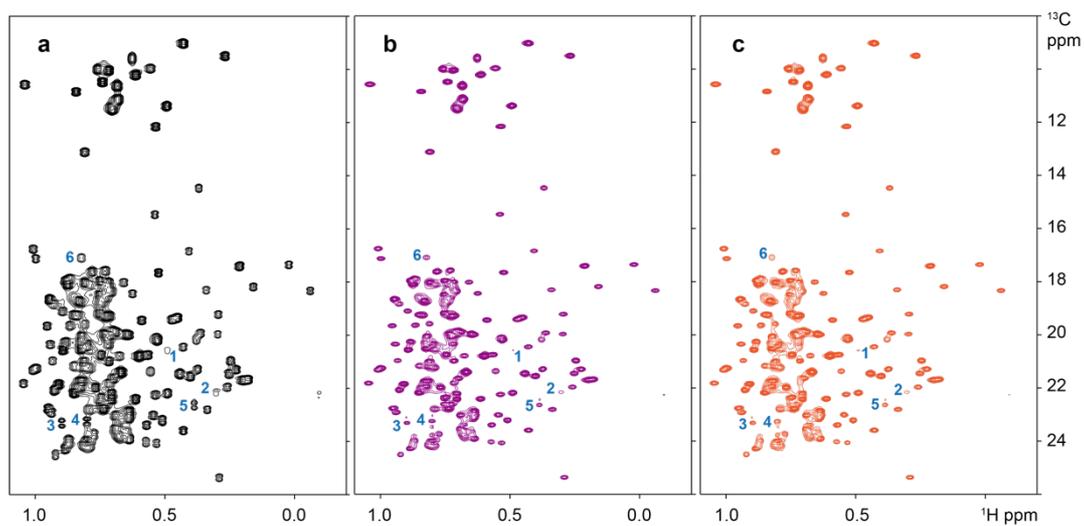

**Fig. S2. Decoupling of the 2D $^1$H-$^{13}$C HMQC spectrum of MALT1 by by the LRD method**. **(a)** is the J-coupled spectrum. **(b)** and **(c)** are decoupled spectra from fully-sampled and 40% NUS, respectively. Although some moderate-intensity peaks are weakened in **(c)** (see for example peak 6), peaks with low intensity such as peaks 1 and 2, are preserved.